\documentclass[prd,superscriptaddress,preprint]{revtex4}

\usepackage{mathrsfs,amsmath}
\usepackage{eurosym}
\usepackage{amssymb}
\usepackage{bbold}

\usepackage{latexsym,epsfig,epstopdf}
\usepackage{amssymb,amsmath,amsthm}

\usepackage{times}
\usepackage{color}

\usepackage{lmodern}
\usepackage[utf8]{inputenc}

\usepackage{graphicx}

\usepackage{color}

\usepackage{braket}

\usepackage[mathscr]{euscript} 

\def\be{\begin{equation}}
\def\ee{\end{equation}}
\def\bea{\begin{eqnarray}}
\def\eea{\end{eqnarray}}

\begin{document}

\title{Klein-Gordon theory in noncommutative phase space}
\author{Shi-Dong Liang}
\email{stslsd@mail.sysu.edu.cn}
\affiliation{School of Physics, Sun Yat-Sen University, Guangzhou 510275, People's
Republic of China,}

\date{\today }

\begin{abstract}
We extend the three-dimensional noncommutative relations of the positions and momenta operators to those in the four dimension. Using the Bopp shift technique, we give the Heisenberg representation of these noncommutative algebras and endow the noncommutative parameters associated with the Planck constant, Planck length and cosmological constant. As an analog with the electromagnetic gauge potential, the noncommutative effect can be interpreted as an effective gauge field, which depends on the Plank constant and cosmological constant. Based on these noncommutative relations, we give the Klein-Gordon (KG) equation and its corresponding current continuity equation in the noncommutative phase space including the canonical and Hamiltonian forms and their novel properties beyond the conventional KG equation. We analyze the symmetries of the KG equations and some observables such as velocity and force of free particles in the noncommutative phase space. We give the perturbation solution of the KG equation.

\end{abstract}


\maketitle
\tableofcontents


\section{Introduction}
The successful applications of quantum mechanics and general relativity imply physics world running noncommutative algebra and noneuclidian geometry in microscopic world and gravity. However, the spacetime background incompatibility between quantum mechanics and general relativity is still a deep mysteries behind nature.\cite{Carlo} This puzzle inspires many attempts to construct a unified theoretical framework for physics world.\cite{Carlo,Thomas}

In particular, understanding physics behind dark energy and dark matter in cosmological observations is a challenge for theoretical physicist.
Even a few possible candidates, such as non-baryonic particle and weakly interacting particle (WIMP), could be
dark matter, while dark energy could be interpreted phenomenologically as cosmological constant, one cannot find any signal of the baryon-type dark matter from the particle physics experiments,\cite{Sabino}. The theoretical prediction of the vacuum energy density for dark energy shows $10^{121}$ times larger than the observed value $10^{-47}$ GeV.\cite{Peebles,Fredenhagen} These results imply that we need new ideas, such as quantization of spacetime, noncommutative geometry and quantum cosmology or quantum gravity to solve these puzzles.\cite{Douglas,Szabo}

On the other hand, the intrinsic singularities in cosmology and particle physics hint us quantizing spacetime and playing
noncommutative algebra.\cite{Konechny,Rosenbaum} The early attempts to quantize spacetime was to assume the noncommutative spacetime emerging in the Planck's scale to remove the singularity of particle physics without renormalizable techniques.\cite{Snyder}
C. N. Yang generalized this idea to curve space to cover gravitational effects.\cite{Yang}

What can we observe from the noncommutative quantum mechanics? Can we detect some signals or effects coming from the Planck's scale and noncommutative spacetime algebra? In condensed matter physics, one finds that the two-dimensional(2D) electronic system in magnetic field is equivalent naturally to the 2D free electronic systems working on the noncommutative phase space.\cite{Delduc,Kovacik,Gamboa}
The noncommutative spacetime is generalized to the noncommutative phase space, namely noncommutative quantum mechanics.\cite{Ho,Gouba,Bellucci} Several schemes were proposed to implement the noncommutative phase space,\cite{Gouba,Mendes} such as the canonical formulation, and  Moyal product, the path integration,\cite{Chaichiian1} the Weyl-Wigner phase space,\cite{Bellucci,Gomes,Ho}  and Seiberg-Witten map with the Bopp shift technique. \cite{Seiberg,Kokado,Bastos}
The noncommutative phase space also extended to the smear Hilbert space to generalize the uncertainty relations. These generalized quantization schemes turn out some novel effects coming from the noncommutative phase space, such as Aharonov-Bohm effect \cite{Das,Liang1,Miguel}, quantum Hall effect\cite{Lapa,Berard}, magnetic monopoles and Berry phase.\cite{Liang2,Harko} However, the noncommutative effects are very difficult to be observed in nano-scale systems because in general it emerges in the Planck's scale $10^{-33}m$. In particular, the noncommutative phase space can be extended to the smeared Hilbert space associated with the generalized and extended uncertainty relations.\cite{Lake1,Lake2}

In general, spacetime can be classified into four regions for different scales, Planck's scale, microscopic and macroscopic scales, and universe scale. Physics in different regions runs different algebras with different energy scales.
The noncommutative spacetime emerges in the Planck and universe scales and the Heisenber's commutative relations works in the microscopic scale. The energy-dependent noncommutative phase space was proposed for understanding novel phenomena in different energy scales\cite{Harko}.

However, most of the previous studies of the noncommutative quantum mechanics are focused on the 2D or 3D phase space. \cite{Delduc,Chaichian2,Kokado,Falomir,Dey}

Interestingly, as a quantum analog, one gave the Lotka-Volterra dynamics in the noncommutative phase space and analyzed its thermodynamic and statistical behaviors.\cite{Bernardini} The quantum dissipation induced in the noncommutative space.\cite{Sivasubramanian}
The Robertson-Schr$\ddot{o}$dinger uncertainty relation was extended the version in the noncommutative phase space.\cite{Bastos1}
Recently, a few attempts devoted to the noncommutative relativistic quantum mechanics and double quantization,\cite{Zou,Gubitosi} including emergence of spin and intrinsic dipole moment in the noncommutative space,\cite{Zou,Calmet,Gomes2} deformed special relativity,\cite{Ghosh} and Snyder model.\cite{Mignemi} More importantly, what physical mechanism hides in the noncommutative phase space?
One expects to extend the 3D noncommutative phase space to the 4D case,
which can provide a foundation of a self-consistent formalism to unify quantum theory and gravitational theory.

In this paper, we propose a 4D noncommutative relations between the 4-vectors of the position and momentum operators in Sec. II. We refer it as to the 4D noncommutative phase space. Based on the Seiberg-Witten map with the Bopp shift technique, we give the Heisenberg's representation of these noncommutative relations.
In Sec III, we give the Klein-Gordon equation and continuity equation in the noncommutative phase space, including the canonical and Hamiltonian forms of the KG equation. We find that the noncommutative effects can be interpreted as an analog with the effective gauge potential.
Interestingly, the free particle carries with an intrinsic velocity, force and acceleration in the noncommutative phase space.
Using the perturbation theory, we obtain the solution of the KG equation and discuss its physical meaning, We also give the non relativistic form of the KG equation and its solution. In Sec. IV, We discuss the basic symmetry of the KG equation, such as the parity, and time reversal symmetries. Finally, in Sec. V, we give the conclusions and outlook.

\section{Noncommutative algebra in Hilbert space}
\subsection{Noncommutative relations of operators}
Let us consider the noncommutative algebra arising in the Planck scale for avoiding singularity of spacetime. We introduce
the 4-vectors of spacetime and momentum operators are defined by
\begin{equation}\label{4VT1}
\widehat{x}^\mu:=\left(
\begin{array}{c}
c\widehat{t} \\
\widehat{x} \\
\widehat{y} \\
\widehat{z}
\end{array}
\right), \quad
\widehat{p}_\mu:=\left(
\begin{array}{c}
\frac{1}{c}\widehat{E} \\
\widehat{p}_x \\
\widehat{p}_y \\
\widehat{p}_z
\end{array}
\right),
\end{equation}
where the operators with hat denote operators in the noncommutative phase space. Thus, the 4D noncommutative relations are defined
\begin{widetext}
\begin{equation}  \label{XXCR1}
\left[\widehat{x}_\mu,\widehat{x}^{\nu}\right]:=\left(
\begin{array}{cccc}
\left[ c\widehat{t},c\widehat{t} \right] & \left[ c\widehat{t},\widehat{x} \right] & \left[ c\widehat{t},\widehat{y} \right] & \left[ c\widehat{t},\widehat{z} \right]\\
\left[ \widehat{x},c\widehat{t} \right] & \left[ \widehat{x},\widehat{x} \right] & \left[ \widehat{x},\widehat{y} \right] & \left[ \widehat{x},\widehat{z} \right]\\
\left[ \widehat{y},c\widehat{t} \right] & \left[ \widehat{y},\widehat{x} \right] & \left[ \widehat{y},\widehat{y} \right] & \left[ \widehat{y},\widehat{z} \right]\\
\left[ \widehat{z},c\widehat{t} \right] & \left[ \widehat{z},\widehat{x} \right] & \left[ \widehat{z},\widehat{y} \right] & \left[ \widehat{z},\widehat{z} \right]
\end{array}
\right):= \Theta^{\mu\nu}
\end{equation}
\end{widetext}

\begin{widetext}
\begin{equation}  \label{PPCR1}
\left[\widehat{p}_\mu,\widehat{p}_{\nu}\right]:=\left(
\begin{array}{cccc}
\left[ \widehat{p}_0,\widehat{p}_0 \right] & \left[ \widehat{p}_0,\widehat{p}_x \right] & \left[ \widehat{p}_0,\widehat{p}_y \right]
& \left[ \widehat{p}_0,\widehat{p}_z \right]\\
\left[ \widehat{p}_x,\widehat{p}_0 \right] & \left[ \widehat{p}_x,\widehat{p}_x \right] & \left[ \widehat{p}_x,\widehat{p}_y \right]
& \left[ \widehat{p}_x,\widehat{p}_z \right]\\
\left[ \widehat{p}_y,\widehat{p}_0 \right] & \left[ \widehat{p}_x,\widehat{p}_x \right] & \left[ \widehat{p}_y,\widehat{p}_y \right]
& \left[ \widehat{p}_y,\widehat{p}_z \right]\\
\left[ \widehat{p}_z,\widehat{p}_0 \right] & \left[ \widehat{p}_z,\widehat{p}_x \right] & \left[ \widehat{p}_z,\widehat{p}_y \right]
& \left[ \widehat{p}_z,\widehat{p}_z \right]
\end{array}
\right):=\Phi_{\mu\nu} 
\end{equation}
\end{widetext}
and
\begin{widetext}
\begin{equation}  \label{XPCR1}
\left[\widehat{x}^\mu,\widehat{p}_{\nu}\right]:=\left(
\begin{array}{cccc}
\left[ c\widehat{t},\widehat{p}_0 \right] & \left[ c\widehat{t},\widehat{p}_x \right] & \left[ c\widehat{t},\widehat{p}_y \right]
& \left[ c\widehat{t},\widehat{p}_z \right]\\
\left[ \widehat{x},\widehat{p}_0 \right] & \left[ \widehat{x},\widehat{p}_x \right] & \left[ \widehat{x},\widehat{p}_y \right]
& \left[ \widehat{x},\widehat{p}_z \right]\\
\left[ \widehat{y},\widehat{p}_0 \right] & \left[ \widehat{y},\widehat{p}_x \right] & \left[ \widehat{y},\widehat{p}_y \right]
& \left[ \widehat{y},\widehat{p}_z \right]\\
\left[ \widehat{z},\widehat{p}_0 \right] & \left[ \widehat{z},\widehat{p}_x \right] & \left[ \widehat{z},\widehat{p}_y \right]
& \left[ \widehat{z},\widehat{p}_z \right]
\end{array}
\right):=\Omega^\mu\ _\nu 
\end{equation}
\end{widetext}
where the matrices of the commutative relations (\ref{XXCR1}), (\ref{PPCR1}) and (\ref{XPCR1}) describe the features of the noncommutative phase space. We will use a set of parameters to characterize the noncommutative strength and there exist some constraints of the parameters when we map the noncommutative algebra to the Heisenberg algebra in the following section. We will endow the parameters with physical meanings in the following sections.

\subsection{Bopp shift and Heisenberg representation}
The noncommutative relations of the position and momentum operators in (\ref{XXCR1}), (\ref{PPCR1}) and (\ref{XPCR1}) give noncommutative operator algebra beyond the canonical (or Heisenberg) algebra in the canonical quantum mechanics.
In general, we have two approaches to implement this algebra. One is to play directly the noncommutative algebra in the noncommutative phase space or based on the Moryal product technique,\cite{Bellucci,Gamboa,Ho,Gouba,Chaichiian1} but it is not easy to reveal the noncommutative effects beyond the canonical quantum mechanics. The other approach is to map this noncommutative algebra to the Heisenberg algebra based on the Seiberg-Witten (SW) map with the Bopp shift technique.\cite{Seiberg,Kokado,Bastos} Here we adopt the Bopp shift to connect the noncommutative algebra to the Heisenberg (or canonical) commutative algebra. \cite{Liang1,Harko,Liang2} Thus, the noncommutative relations (\ref{XXCR1}), (\ref{PPCR1}) and (\ref{XPCR1}) can be expressed in terms of the Heisenberg commutative relations with additional effects from the noncommutative phase space.\cite{Liang1, Harko, Liang2}

Let us define $\widehat{x}^\mu, \widehat{p}_\mu\in \widehat{\mathcal{O}}_{nc}$ living in the noncommutative phase space, where $\widehat{x}^\mu:=\left(c\widehat{t}, \widehat{\mathbf{x}}\right)\equiv\left(c\widehat{t},\widehat{x},\widehat{y},\widehat{z}\right)$, and $\widehat{p}_\mu:=\left(\widehat{p}_0,\widehat{\mathbf{p}}\right)\equiv\left(\widehat{p}_0,\widehat{p}_x,\widehat{p}_y,\widehat{p}_z\right)$,
and define $x^\mu, p_\mu\in \mathcal{O}_H$ living in the Heisenberg phase space, where $x_\mu:=\left(ct,\mathbf{x}\right)\equiv\left(ct,x,y,z\right)$, and $p_\mu:=\left(p_0,\mathbf{p}\right)\equiv\left(ct,p_x,p_y,p_z\right)$.

The Seiberg-Witten (SW) map with the Bopp shift is defined by
\begin{equation}\label{SWM1}
\mathcal{M}_{SW}: \widehat{\mathcal{O}}_{nc}\ni
\begin{array}{ccc}
\left(\widehat{x}^\mu,\widehat{p}_\mu\right) & \rightarrow & \left(x^\mu,p_\mu\right), \\
\widehat{O}\left(\widehat{x}^\mu,\widehat{p}_\mu\right) & \rightarrow & O\left(x^\mu,p_\mu\right),
\end{array} \in \mathcal{O}_H
\end{equation}
where $\widehat{O}\left(\widehat{x}^\mu,\widehat{p}_\mu\right)$ means any operator expressed in terms of $\left(\widehat{x}^\mu,\widehat{p}_\mu\right)$ in the noncommutative phase space, such as momentum and Hamiltonian operators. They obey the noncommutative relations in (\ref{XXCR1}), (\ref{PPCR1}) and (\ref{XPCR1}), while $O\left(x^\mu,p_\mu\right)$ represents any operator expressed in terms of  $\left(x^\mu,p_\mu\right)$. They obey the Heisenberg commutative relations,
\begin{subequations}\label{HSBCR1}
\begin{eqnarray}
\left[x^\mu,x^\nu\right] &=& 0, \\
\left[p_\mu,p_\nu\right] &=& 0, \\
\left[x^\mu,p_\nu\right] &=& i\hbar\delta^\mu\ _\nu.
\end{eqnarray}
\end{subequations}
Thus, we can obtain a claim in the following.

\textbf{Claim I}: The Bopp shift in (\ref{SWM1}) is constructed by
\begin{subequations} \label{NCXP2}
\begin{eqnarray}
\left(
\begin{array}{c}
c\widehat{t} \\
\widehat{x} \\
\widehat{y} \\
\widehat{z}
\end{array}
\right)&:=& \left(
\begin{array}{c}
ct-\frac{\theta }{\hbar}\left(p_x+p_y+p_z\right) \\
x -\frac{\theta}{2\hbar}\left(p_y+p_z\right) \\
y +\frac{\theta}{2\hbar}\left(p_x-p_z\right) \\
z +\frac{\theta}{2\hbar}\left(p_x+p_y\right)
\end{array}
\right) \\
\left(
\begin{array}{c}
\widehat{p}^0 \\
\widehat{p}^{x} \\
\widehat{p}^{y} \\
\widehat{p}^{z}
\end{array}
\right) &:=& \left(
\begin{array}{c}
\frac{i\hbar}{c} \frac{\partial}{\partial t}+\frac{\eta}{\hbar}(x+y+z) \\
p_{x} +\frac{\eta}{2\hbar}(y+z) \\
p_{y} -\frac{\eta}{2\hbar}(x-z) \\
p_{z} -\frac{\eta}{2\hbar}(x+y)
\end{array}%
\right)
\end{eqnarray}
\end{subequations}
such that the noncommutative algebra of operators in (\ref{XXCR1}), (\ref{PPCR1}) and (\ref{XPCR1})
in the noncommutative phase space can be implemented equivalently by the Heisenberg commutative algebra (\ref{HSBCR1}) in the canonical phase space.

\textbf{Proof}: The Bopp shift (\ref{NCXP2}) can be rewritten as a matrix form
\begin{subequations}\label{SWXP1}
\begin{eqnarray}
\widehat{x}^\mu   &=& x^\mu+\lambda^{\mu\nu} p_\nu \\
\widehat{p}_\mu   &=& p_\mu+\pi_{\mu\nu} x^\nu,
\end{eqnarray}
\end{subequations}
where
\begin{equation}\label{LDa1}
\left[\lambda^{\mu\nu}\right]=\left(
\begin{array}{cccc}
0 & -\frac{\theta}{\hbar}& -\frac{\theta}{\hbar} & -\frac{\theta}{\hbar} \\
0 & 0 &  -\frac{\theta}{2\hbar}& -\frac{\theta}{2\hbar} \\
0 & \frac{\theta}{2\hbar} & 0 & -\frac{\theta}{2\hbar} \\
0 & \frac{\theta}{2\hbar} & \frac{\theta}{3\hbar} & 0
\end{array}\right)
\end{equation}
and
\begin{equation}\label{PI1}
\left[\pi^{\mu\nu}\right]=\left(
\begin{array}{cccc}
0 & \frac{\eta}{\hbar} & \frac{\eta}{\hbar} & \frac{\eta}{\hbar}\\
0 & 0 &  \frac{\eta}{2\hbar}& \frac{\eta}{2\hbar} \\
0 & -\frac{\eta}{2\hbar} & 0 & \frac{\eta}{2\hbar} \\
0 & -\frac{\eta}{2\hbar} & -\frac{\eta}{2\hbar} & 0
\end{array}\right)
\end{equation}

Using the Heosenberg commutative relations in (\ref{HSBCR1}), the noncommutative relations between the position operators are expressed as
\begin{eqnarray}\label{XXNCR0}
\left[\widehat{x}^\mu,\widehat{x}^{\nu}\right] &=& \left[x^\mu+\lambda^{\mu\kappa} p_\kappa, \left(x^\nu+\lambda^{\nu\sigma}p_\sigma\right) \right] \nonumber \\
   &=& \left[x^\mu, p_{\sigma} \right]\lambda^{\sigma\nu} + \lambda^{\nu\kappa}\left[ p_\kappa, x^\nu \right] \nonumber\\
   &=& i\hbar\left(\lambda^{\mu\nu}- \lambda^{\nu\mu}\right)\equiv i\Theta^{\mu\nu},
\end{eqnarray}
where $\lambda^{\mu\nu}$ is the transposition of $\lambda^{\nu\mu}$.
Playing the matrix subtraction in (\ref{PI1}), the noncommutative relation is obtained
\begin{equation}\label{XXNCR1}
\left[\widehat{x}^\mu,\widehat{x}^{\nu}\right] =i\Theta_{\mu\nu},
\end{equation}
where
\begin{equation}\label{XXNCR2}
\left[\Theta_{\mu\nu}\right]=\left(
\begin{array}{cccc}
0 & \theta & \theta & \theta \\
-\theta & 0 & \theta & \theta \\
-\theta & -\theta & 0 & \theta \\
-\theta & -\theta & -\theta & 0
\end{array}
\right).
\end{equation}
This is the noncommutative relation (\ref{XXCR1}).

Similarly, the noncommutative relations between the momentum operators are expressed as
\begin{eqnarray}
\left[\widehat{p}_\mu,\widehat{p}_{\nu}\right] &=& \left[p_\mu+\pi_{\mu\kappa} x^\kappa, \left(p_\nu+\pi_{\nu\sigma}x^\sigma\right) \right] \nonumber\\
   &=& \left[p_\mu, x^\sigma \right]\pi^{\sigma\nu} +\pi^{\nu\kappa}\left[ x^\kappa, p_\nu \right] \nonumber\\
   &=& i\hbar \left(-\pi^{\mu\nu}+\pi^{\nu\mu}\right)\equiv i\Phi_{\mu\nu}.
\end{eqnarray}
By the matrix subtraction of $\pi$, the noncommutative relations are obtained
\begin{equation}\label{PPNCR1}
\left[\widehat{p}_\mu,\widehat{p}_{\nu}\right]=i\Phi_{\mu\nu},
\end{equation}
where
\begin{equation}\label{PPNCR2}
\left[\Phi_{\mu\nu} \right]=\left(
\begin{array}{cccc}
0 & \eta & \eta & \eta \\
-\eta & 0 & \eta & \eta \\
-\eta & -\eta & 0 & \eta \\
-\eta & -\eta & -\eta & 0
\end{array}\right).
\end{equation}
This is the noncommutative relation (\ref{PPCR1}).

In the same way, the noncommutative relations between the position and momentum operators are expressed as
\begin{eqnarray}
\left[\widehat{x}^\mu,\widehat{p}_{\nu}\right] &=& \left[x^\mu+\lambda^{\mu\kappa} p_\kappa, \left(p_\nu+\pi_{\nu\sigma}x^\sigma\right) \right] \nonumber\\
   &=& \left[x^\mu, p_\nu \right] + \lambda^{\mu\kappa} \left[ p_\kappa, x^{\sigma}\right]\pi_{\sigma\nu} \nonumber\\
   &=& i\hbar \left(\delta^\mu\ _\nu-\lambda^{\mu\kappa} \pi_{\kappa\nu}\right)\nonumber \\
   &\equiv & i\Omega^\mu\ _\nu.
\end{eqnarray}
Making the matrix product of $\lambda$ and $\pi$, the noncommutative relation can be obtained
\begin{equation}\label{XPNCR1}
\left[\widehat{x}^\mu,\widehat{p}_{\nu}\right]=i\Omega^\mu\ _\nu,
\end{equation}
where
\begin{equation}\label{XPNCR2}
\left[\Omega^\mu\ _\nu\right] =\left(
\begin{array}{cccc}
\hbar + 3\frac{\theta\eta}{\hbar} & \frac{\theta\eta}{\hbar}  & 0 & -\frac{\theta\eta}{\hbar}\\
\frac{\eta\theta}{\hbar} & \hbar+ \frac{\theta\eta}{2\hbar} &   \frac{\theta\eta}{4\hbar} &  -\frac{\theta\eta}{4\hbar} \\
0 &  \frac{\theta\eta}{4\hbar} & \hbar+ \frac{\theta\eta}{2\hbar} &  \frac{\theta\eta}{4\hbar} \\
-\frac{\eta\theta}{\hbar}  & -\frac{\theta\eta}{4\hbar} &  \frac{\theta\eta}{4\hbar} & \hbar+ \frac{\theta\eta}{2\hbar}
\end{array}\right).
\end{equation}

This representation of the position and momentum operators can be regarded as a Heisenbereg representation of the noncommutative quantum mechanics. The Bopp shift provides an efficient way to do the Heisenberg algebra in the noncommutative phase space even though the Bopp shift is not unitary and canonical. It can be verified that the commutative relations (\ref{XXNCR1}), (\ref{XPNCR1}) and (\ref{PPNCR1}) satisfy the Lorentz invariants.

\subsection{What physics involves noncommutative algebra?}
What are the physical meanings of the noncommutative parameters? In the canonical quantum mechanics, the Planck constant plays a role to quantize the intrinsic physical space of elementary particles, such that the quantum state is described by the Hilbert space. To explore physics in the noncommutative phase space, we endow the noncommutative parameters with physical meaning. Based on the Bopp shift, there are only two independent parameters, $\theta$ and $\eta$. In general, physics involved the noncommutative relations happen in the Planck and universe scales, such as intrinsic spacetime singularities, dark matter and dark energy in universe. In particular, the interaction between photon and gravity implies that there exists a minimum length of spacetime in the Planck scale. The dark energy in cosmology can be regarded as an intrinsic minimum curvature of spacetime, which can be interpreted phenomenally as cosmological constant. Thus, we propose a parameterization scheme of the noncommutative parameters associated with a set of physical constants,

\begin{equation}\label{PMTT}
\theta=\ell_{P}^2;  \quad \eta=\hbar^2 \Lambda,
\end{equation}
where $\ell_{P}=\sqrt{\frac{\hbar G}{c^3}}$ is the Planck length, and $\hbar$ is the Planck constant. $G$ is the gravitational constant.
$\Lambda\simeq 10^{-56}cm^{-2}$ is the cosmological constant. Based on this parameter setting, we obtain the matrices of the noncommutative relations,
\begin{equation}\label{XXNCR3}
\left[\Theta_{\mu\nu}\right]=\left(
\begin{array}{cccc}
0 & \ell_P^2 & \ell_P^2 & \ell_P^2\\
-\ell_P^2 & 0 &  \ell_P^2 &  \ell_P^2 \\
-\ell_P^2 & - \ell_P^2 & 0 &  \ell_P^2 \\
-\ell_P^2 & - \ell_P^2 & - \ell_P^2 & 0
\end{array}\right).
\end{equation}

\begin{equation}\label{PPNCR3}
\left[\Phi_{\mu\nu}\right]=\hbar\left(
\begin{array}{cccc}
0 & \hbar \Lambda  & \hbar \Lambda & \hbar \Lambda \\
-\hbar \Lambda & 0 &  \hbar\Lambda &  \hbar\Lambda \\
-\hbar \Lambda &  -\hbar\Lambda & 0 &  \hbar\Lambda \\
-\hbar \Lambda & -\hbar\Lambda &  -\hbar\Lambda & 0
\end{array}\right).
\end{equation}

\begin{widetext}
\begin{equation}\label{XPNCR3}
\left[\Omega_{\mu\nu} \right]=\left(
\begin{array}{cccc}
\hbar + 3\hbar  \Lambda \ell_P^2 & \hbar \Lambda \ell_P^2  & 0 & -\hbar \Lambda \ell_P^2\\
\hbar \Lambda \ell_P^2 & \hbar+ \frac{\hbar\ell_P^2 \Lambda}{2} &   \frac{\hbar\ell_P^2\Lambda}{4} &  -\frac{\hbar\ell_P^2\Lambda}{4} \\
0 &  \frac{\hbar\ell_P^2\Lambda}{4} & \hbar+ \frac{\hbar\ell_P^2\Lambda}{2} &  \frac{\hbar\ell_P^2\Lambda}{4} \\
-\hbar \Lambda \ell_P^2  & -\frac{\hbar\ell_P^2\Lambda}{4} &  \frac{\hbar\ell_P^2\Lambda}{4} & \hbar+ \frac{\hbar\ell_P^2\Lambda}{2}
\end{array}\right).
\end{equation}
\end{widetext}

It should be remarked that in general, how to endow the parameters with physics meanings depends on what problem we solve. The parameter setting (\ref{PMTT}) is based on physics we concern in Planck and universe scales. In principle, we can adopt other settings when we concern microscopic or macroscopic physics such as the de Broglie wave length, which depends on the particle momentum or energy $\lambda_d=\frac{\hbar}{p}$. Thus, the de Broglie wave length as a variable extend the formulation to microscopic regions. In fact, the energy-dependent noncommutative quantum mechanics has be proposed for some applications.\cite{Harko}

\section{Klein-Gordon equation}
\subsection{Canonical form of Klein-Gordon equation}
Suppose that the energy-momentum relation in the noncommutative phase space still hold, the Klein-Gordon equation  is given by

\begin{equation}\label{KGE1}
\left(\widehat{p}_\mu \widehat{p}^\mu-m_0^2c^2\right)\phi=0,
\end{equation}
where the momentum operators obey the noncommutative relations (\ref{PPCR1}) and $\phi\equiv\phi(x^\mu)$. Hereafter we omit the variables $x^\mu$ for convenience.
To explore what novel properties coming from the noncommutative phase space, we transform the KG equation to the Heisenberg's representation
based on the SW map (\ref{SWM1}).
Note that the analog of electrodynamic field with the minimum coupling, the momenta in the noncommutative phase space can be rewritten as
\begin{equation}\label{MCP1}
\widehat{p}_\mu=p_\mu-A_\mu,
\end{equation}
where $A_\mu$ can be viewed as an effective gauge potential induced by the noncommutative phase space,
\begin{equation}\label{AAA1}
A_\mu:=-\pi_{\mu\nu} x^\nu=\left(
\begin{array}{c}
-\frac{\eta}{\hbar} (x+y+z) \\
-\frac{\eta}{2\hbar} (y+z) \\
\frac{\eta}{2\hbar} (x-z) \\
\frac{\eta}{2\hbar} (x+y)
\end{array}\right).
\end{equation}
Note that $\widehat{p}^\mu=g^{\mu\nu}\widehat{p}_\nu $, where $g^{\mu\nu}$ is the Lorentz metric. The signature here is $(+, -, -, -)$. Thus, the Klein-Gordon (KG) equation can be rewritten as
\begin{equation}\label{KGE1}
\left[\left(p_\mu -A_\mu\right)\left(p^\mu-A^\mu\right)-m_0^2c^2\right]\phi=0.
\end{equation}
where
\begin{equation}\label{MCPS1}
p_\mu=i\hbar\left(\frac{1}{c}\frac{\partial}{\partial t},\nabla\right)\equiv i\hbar\partial_\mu
\end{equation}
is the 4-vector momentum in the Heisenberg's canonical representation. In other words, the operators $x_\mu$ and $p_\mu$ play the Heisenberg's commutative relations. The gauge potential $A_\mu$ descries some novel effects from the noncommutative phase space.
\begin{equation}\label{AAA2}
A_\mu=\left(A_0 , \mathbf{A}\right)
\end{equation}
is the 4-vector potential (or gauge potential) generated from the noncommutative phase space, in which
\begin{subequations}\label{AA2}
\begin{eqnarray}
A_0 &=& -\frac{\eta}{\hbar}  (x+y+z) \\
\mathbf{A} &=& \frac{\eta}{2\hbar}\left[(y+z)\mathbf{i}-(x-z)\mathbf{j}-(x+y)\mathbf{k}\right]
\end{eqnarray}
\end{subequations}
The gauge transformation is defined by
\begin{equation}\label{GT1}
A_\mu\rightarrow A'_\mu=A_\mu-\frac{\partial \chi}{\partial x_\mu}, \quad \phi\rightarrow \phi'=e^{i\Phi}\phi.
\end{equation}
More explicitly, the gauge transformation is expressed as
\begin{subequations}
\begin{eqnarray}
A_0 & \rightarrow & A'_0-\frac{1}{c}\frac{\partial \chi}{\partial t} \\
\mathbf{A} & \rightarrow & \mathbf{A}'+\nabla \chi, \\
\phi &\rightarrow &\phi' = e^{i\Phi}\phi, \quad \Phi=\frac{\chi}{\hbar }.
\end{eqnarray}
\end{subequations}
It can be verified that the KG equation (\ref{KGE1}) is gauge invariants and Lorentz covariants.
Interestingly, the gauge field coupled with the KG equation does not contain a coupled constant. In other words, the coupled constant is dimensionless.

By expending the product in the KG equation (\ref{KGE1}) and using (\ref{AAA1}) and (\ref{MCPS1}), $\partial_\mu A^\mu=0$ for $\partial_t x^i=0$. The KG equation can be rewritten as a tensorial form,
\begin{equation}\label{KGTeq1}
\left[\partial_\mu\partial^\mu-i\frac{2}{\hbar}A_\mu \partial^\mu+\frac{1}{\hbar^2}A_\mu A^\mu +\frac{m_0 c^2}{\hbar^2}\right]\phi=0
\end{equation}
where $\partial_\mu\partial^\mu:=\frac{1}{c^2}\partial_t^2-\nabla^2$ is the d'Alembert operator. Since all terms in the KG equation are Lorentz scaler, the KG equation in the noncommutative phase space is Lorentz covariant.

\subsection{Noncommutative algebra, gauge field and cosmological constant}

As an analog with electrodynamics, we define the gauge field generated from the gauge potential in the noncommutative phase space,
\begin{equation}\label{GF1}
F_{\mu\nu}=\partial_\mu A_\nu-\partial_\nu A_\mu
\end{equation}
For $\partial_t x^i=0$, using the parameterization scheme (\ref{PMTT}) and plugging the gauge potential (\ref{AAA1}) into (\ref{GF1}), the gauge field is obtained
\begin{equation}\label{GF2}
\left[F_{\mu\nu}\right]=\left(
\begin{array}{cccc}
0 & \hbar\Lambda & \hbar\Lambda  & \hbar\Lambda \\
-\hbar\Lambda  & 0 & \hbar\Lambda & \hbar\Lambda\\
-\hbar\Lambda  & -\hbar\Lambda & 0 & \hbar\Lambda\\
-\hbar\Lambda  & -\hbar\Lambda & -\hbar\Lambda & 0
\end{array}
\right).
\end{equation}
Interestingly, the effective gauge field depends on the Planck constant and cosmological constant, which is consistent with the interpretation of the dark energy emerged from the cosmological constant. There does not exist a coupled constant between the effective gauge field and dynamical equation. This is quite different from conventional electromagnetic $U(1)$ gauge field and Yang-Mills $SU(2)$ gauge field. Moreover, there exists a natural gauge $\partial_\mu A^\mu=0$. These features of the effective gauge field show an intrinsic geometry (curvature) of spacetime coupled with the dynamical equation. In other words, this property can be interpreted as a natural coupling between quantum and gravity.

\subsection{Probability current and continuity equation}
To give the probability current and continuity equation, following the similar steps of those for the KG equation coupled with electromagnetic field, making the product of the complex conjugated wave function $\phi^*$ in the left-hand side of (\ref{KGE1}), subsequently subtracting its complex conjugate, we can obtain the continuity equation,\cite{Armin}
\begin{equation}\label{CEQ1}
\partial_\mu J^\mu=0,
\end{equation}
where $J^\mu=(c\rho,\mathbf{j})$ is the 4-vector probability current density. This tensorial form of the current continuity equation implies that it is Lorentz invariant. The probability and current densities are given respectively by
\begin{subequations}\label{RoJ}
\begin{eqnarray}
\rho &=& \frac{i\hbar}{2m_0c^2}\left(\phi^* \frac{\partial \phi}{\partial t}- \frac{\partial \phi^*}{\partial t}\phi\right) -\frac{1}{m_0c}A^0\phi^*\phi\\
\mathbf{j} &=& -\frac{i\hbar}{2m_0}\left[\phi^* \nabla \phi- (\nabla \phi^*)\phi\right] -\frac{1}{m_0}\mathbf{A}\phi^*\phi,
\end{eqnarray}
\end{subequations}
which are analog with the KG equation coupled with electromagnetic field. The tensorial form of the continuity equation (\ref{CEQ1}) means that
the current continuity equation is Lorentz covariant.
However, $\rho$ is not positive definite because $\frac{\partial \phi}{\partial t}$ is arbitrary for the second-order differential equation. Hence, $\rho$ cannot be interpreted strictly as the probability density and thus $\mathbf{j}$ also cannot be defined strictly as the density of the probability current.
Interestingly both $\rho$ and $\mathbf{j}$ depend on the effective gauge potential induced in the noncommutative phase space.
What effect of the gauge potential plays in the noncommutative phase space?
We will examine the KG equation by the perturbation theory in the following section.

\subsection{Hamiltonian form of Klein-Gordon equation}

To reveal more physics of the KG equation, we can convert the canonical form of the KG equation to its Hamiltonian form.\cite{Armin}
Let us define
\begin{subequations}\label{Pichi1}
\begin{eqnarray}
\phi &:=&\varphi+\chi \\
\left(i\hbar \partial_t-A_0\right)\phi &:=&  m_0c^2(\varphi-\chi),
\end{eqnarray}
\end{subequations}
such that we have
\begin{widetext}
\begin{subequations}\label{Pichi2}
\begin{eqnarray}
m_0c^2\phi+\left(i\hbar \partial_t-A_0\right)\phi &=& m_0c^2(\varphi+\chi) +m_0c^2(\varphi-\chi)=2m_0c^2\varphi \\
m_0c^2\phi-\left(i\hbar \partial_t-A_0\right)\phi &=& m_0c^2(\varphi+\chi)-m_0c^2(\varphi-\chi)=2m_0c^2\chi
\end{eqnarray}
\end{subequations}
\end{widetext}
Thus, solving (\ref{Pichi2}) to express $\varphi$ and $\chi$ in terms of $\phi$, we obtain
\begin{subequations}\label{Pichi3}
\begin{eqnarray}
\varphi &=&\frac{1}{2m_0c^2}\left(m_0c^2+i\hbar \partial_t-A_0\right)\phi \\
\chi &=&\frac{1}{2m_0c^2}\left(m_0c^2-i\hbar \partial_t+A_0\right)\phi
\end{eqnarray}
\end{subequations}
Using (\ref{Pichi1}b), we have
\begin{equation}\label{Pichi4}
\left(i\hbar \partial_t-A_0\right)(\varphi-\chi)=\frac{1}{m_0c^2}\left(i\hbar \partial_t-A_0\right)^2\phi.
\end{equation}
Note that

\begin{equation}\label{PP1}
\left(p^\mu -A^\mu\right)\left(p_\mu-A_\mu\right)=
\left(\frac{i\hbar}{c} \frac{\partial}{\partial t}  -A_0\right)^2-\left(\frac{\hbar}{i}\nabla -\mathbf{A}\right)^2
\end{equation}

and using the canonical form of the KG equation (\ref{KGE1}), the left-hand side of (\ref{Pichi4}) can be rewritten as
\begin{equation}\label{Pichi5}
\left(i\hbar \partial_t-A_0\right)(\varphi-\chi)=\left[\frac{1}{m_0}\left(\frac{\hbar}{i}\nabla -\mathbf{A}\right)^2
+m_0c^2\right]\phi.
\end{equation}
Combining (\ref{Pichi1}) and (\ref{Pichi5}), we have
\begin{widetext}
\begin{subequations}\label{Hf1}
\begin{eqnarray}
\left(i\hbar \partial_t-A_0\right)(\varphi+\chi) &=&  m_0c^2(\varphi-\chi),\\
\left(i\hbar \partial_t-A_0\right)(\varphi-\chi) &=& \left[\frac{1}{m_0}\left(\frac{\hbar}{i}\nabla -\mathbf{A}\right)^2
+m_0c^2\right](\varphi+\chi).
\end{eqnarray}%
\end{subequations}
\end{widetext}
Adding and subtracting (\ref{Hf1}), we obtain
\begin{widetext}
\begin{subequations}\label{Hf2}
\begin{eqnarray}
\left(i\hbar \partial_t-A_0\right)\varphi &=&  \frac{1}{2m_0}\left(\frac{\hbar}{i}\nabla -\mathbf{A}\right)^2(\varphi+\chi)
+2m_0c^2\varphi,\\
\left(i\hbar \partial_t-A_0\right)\chi &=&  -\frac{1}{2m_0}\left(\frac{\hbar}{i}\nabla -\mathbf{A}\right)^2(\varphi+\chi)
-2m_0c^2\chi.
\end{eqnarray}
\end{subequations}
\end{widetext}
By rearranging the terms of $\varphi$ and $\chi$, the equation (\ref{Hf2}) can be rewritten as
\begin{widetext}
\begin{equation}\label{Hf3}
i\hbar \frac{\partial}{\partial t}\left(
\begin{array}{c}
\varphi  \\
\chi
\end{array}\right)
=\left(
\begin{array}{cc}
\frac{1}{2m_0}\left(\frac{\hbar}{i}\nabla -\mathbf{A}\right)^2+m_0c^2+A_0 & \frac{1}{2m_0}\left(\frac{\hbar}{i}\nabla -\mathbf{A}\right)^2   \\
-\frac{1}{2m_0}\left(\frac{\hbar}{i}\nabla -\mathbf{A}\right)^2 &
-\frac{1}{2m_0}\left(\frac{\hbar}{i}\nabla -\mathbf{A}\right)^2-m_0c^2+A_0
\end{array}\right)
\left(
\begin{array}{c}
\varphi  \\
\chi
\end{array}\right)
\end{equation}
\end{widetext}
This is the Hamiltonian form of the KG equation. Let us define $\psi^T:=\left(\varphi, \chi\right)$, the KG equation (\ref{Hf3}) can be rewritten as the Schr\"{o}dinger-like equation,
\begin{equation}\label{SChE}
i\hbar \frac{\partial \psi}{\partial t} =H\psi,
\end{equation}
where the Hamiltonian is given by
\begin{equation}\label{HHH1}
H=\frac{\tau_3+i\tau_2}{2m_0}\left(\frac{\hbar}{i}\nabla -\mathbf{A}\right)^2+\tau_3 m_0c^2+A_0
\end{equation}
where $\tau_{2(3)}$ are the Pauli's matrices. It should be remarked that the Hamiltonian in (\ref{HHH1}) is not Hermitian because $(i\tau_2)^\dagger\neq i\tau_2$.
We will discuss this issue in the next section.

\subsection{Probability and current densities}

Similarly, as an analog with Schr\"{o}dinger equation, the probability density is given by
\begin{equation}\label{CD2}
\rho=\psi^\dagger\tau_3\psi=\varphi^*\varphi-\chi^*\chi,
\end{equation}
and the probability current density is obtained
\begin{widetext}
\begin{equation}\label{CD3}
\mathbf{j}=-\frac{i\hbar}{2m_0}\left[\psi^\dagger\tau_3(\tau_3+i\tau_2)\nabla\psi-\nabla\psi^\dagger\tau_3(\tau_3+i\tau_2)\psi\right]
-\frac{1}{m_0}\mathbf{A}\psi^\dagger\tau_3(\tau_3+i\tau_2)\psi.
\end{equation}
\end{widetext}
However, $\rho$ is also not positive definite such that
$\rho$ cannot be endowed strictly as the probability density
and thus $\mathbf{j}$ also cannot be defined strictly as the current densities.\cite{Armin}

It should be remarked that the noncommutative effects coming from the noncommutative algebra play an effective gauge potential as an analog with the electromagnetic gauge potential, which modifies the densities of the probability and probability current, the velocity and force. However, strictly speaking, the single particle picture is not well-defined because the Hamiltonian is not Hermitian. One proposes to redefine the inner product for Hermitization of operators, which is called the generalized inner product. The detail discussion on this issue may be found in the book. \cite{Armin}

\subsection{Velocity and force}
To examine the basic properties of the relativistic particle in the noncommutative phase space, we consider a free particle model to investigate some dynamical behaviors. Based on the Heisenberg equation (or called as Heisenberg picture), we have two approaches to define  observables. One is that the observables are defined in the noncommutative phase space. We refer it as to Case I, in which
the commutative relations are calculated directly by the noncommutative relations in (\ref{XXCR1}), (\ref{PPCR1}) and (\ref{XPCR1}), and then using the SW map to transform the results to the Heisenberg representation. The other is that the observables are defined in the Heisenberg phase space. We refer it as to Case II, in which the commutative relations are calculated by the Heisenberg relations of the SW map. We will discuss these two cases.

Case I: In the noncommutative phase space, the velocity of particle is defined by $\mathbf{\widehat{v}}=\frac{d\mathbf{\widehat{x}}}{dt}$ in the noncommutative phase space. The Hamiltonian in (\ref{HHH1}) of free particle is given by
\begin{equation}\label{HHH2}
\widehat{H}=\frac{\tau_3+i\tau_2}{2m_0}\widehat{\mathbf{p}}^2+\tau_3 m_0c^2.
\end{equation}
Thus, the velocity of particle is given by the Heisenberg equation
\begin{equation}\label{VP1}
\left\langle\mathbf{\widehat{v}}\right\rangle=\frac{1}{i\hbar}\left\langle\left[\mathbf{\widehat{x}},\widehat{H}\right]\right\rangle,
\end{equation}
where $\langle \widehat{O}\rangle$ means the expectation value of $\widehat{O}$ in state $\psi$.
Using the commutative relations (\ref{XPNCR1}),(see the Appendix A), we have
\begin{equation}\label{2P3}
\left[ \widehat{\mathbf{x}},\widehat{\mathbf{p}}^{2} \right]= 2i\left(\kappa^{\alpha}\widehat{\mathbf{p}}+\kappa^{\beta}\widehat{\mathbf{k}}_v\right),
\end{equation}
where
\begin{equation}\label{2Pr}
\widehat{\mathbf{k}}_v=\left(\widehat{p}_y-\widehat{p}_z\right)\mathbf{i}+\left(\widehat{p}_x+\widehat{p}_z\right)\mathbf{j}
+\left(-\widehat{p}_x+\widehat{p}_y\right)\mathbf{k}.
\end{equation}
Thus, putting the Hamiltonian (\ref{HHH2}) into (\ref{VP1}), the velocity is obtained as
\begin{equation}\label{VP2}
\left\langle\mathbf{\widehat{v}}\right\rangle=\frac{\tau_3+i\tau_2}{m_0\hbar}
\left(\kappa^\alpha\left\langle\mathbf{\widehat{p}}\right\rangle+\kappa^\beta\left\langle\widehat{\mathbf{k}}_v\right\rangle\right).
\end{equation}

Similarly, the force is defined by $\mathbf{\widehat{f}}=\frac{d\widehat{\mathbf{p}}}{dt}$. In the same way, we have
\begin{equation}\label{FC1}
\mathbf{\widehat{f}}=\frac{1}{i\hbar}\left[\widehat{\mathbf{p}},\widehat{H}\right].
\end{equation}
Using the commutative relations (\ref{PPNCR1}), (see the Appendix A), we have
\begin{equation}\label{PP3}
\left[ \widehat{\mathbf{p}},\widehat{\mathbf{p}}^{2} \right]= 2\eta\left[\left(\widehat{p}_y+\widehat{p}_z\right)\mathbf{i}
+\left(-\widehat{p}_x+\widehat{p}_z\right)\mathbf{j}-\left(\widehat{p}_x+\widehat{p}_y\right)\mathbf{k}\right]
\end{equation}
Consequently, the force is obtained as
\begin{widetext}
\begin{equation}\label{ff2}
\left\langle\mathbf{\widehat{f}}\right\rangle=\frac{\tau_3+i\tau_2}{m_0\hbar}\eta\left[\left\langle\widehat{p}_y+\widehat{p}_z)\right\rangle\mathbf{i}
   +\left\langle-\widehat{p}_x+\widehat{p}_z\right\rangle\mathbf{j}-\left\langle\widehat{p}_x+\widehat{p}_y\right\rangle\mathbf{k}\right].
\end{equation}
\end{widetext}

Case II: The velocity of particle is defined by $\mathbf{v}=\frac{d\mathbf{x}}{dt}$ in the Heisenberg phase space.
based on the SW map, the Hamiltonian in (\ref{HHH1}) of free particle is given by
\begin{equation}\label{HHH3}
H=\frac{\tau_3+i\tau_2}{2m_0}(\mathbf{p}-\mathbf{A})^2+\tau_3 m_0c^2+A_0.
\end{equation}
The velocity of particle is given by
\begin{equation}\label{VP3}
\left\langle\mathbf{v}\right\rangle=\frac{d\mathbf{x}}{dt}=\frac{1}{i\hbar}\left\langle\left[\mathbf{x},H\right]\right\rangle.
\end{equation}
Note that $[\mathbf{x},\mathbf{p}^2]=2i\hbar \mathbf{p}$ and $[\mathbf{x},\mathbf{A}]=0$, the velocity is obtained as
\begin{equation}\label{VP4}
\left\langle\mathbf{v}\right\rangle = \frac{\tau_3+i\tau_2}{m_0}\left\langle\mathbf{p}-\mathbf{A}\right\rangle.
\end{equation}

The force is defined by $\mathbf{f}=\frac{d\mathbf{p}}{dt}$ in the Heisenberg phase space. Using the Heisenberg equation, we have
\begin{equation}\label{FC2}
\left\langle\mathbf{f}\right\rangle=\frac{1}{i\hbar}\left\langle\left[\mathbf{p},H\right]\right\rangle.
\end{equation}

Note that $\left[\frac{\partial}{\partial x_\mu}, A_\nu\right]=\frac{\partial A_\nu}{\partial x_\mu}$ and $\frac{\partial A_\mu}{\partial x_\mu}=0$, we have
\begin{subequations}\label{pp2}
\begin{eqnarray}
\left[p_x,(\mathbf{p}-\mathbf{A})^2\right] &=& i\eta (p_y-A_y+p_z-A_z), \\
\left[p_y,(\mathbf{p}-\mathbf{A})^2\right] &=& i\eta (-p_x+A_x+p_z-A_z), \\
\left[p_z,(\mathbf{p}-\mathbf{A})^2\right] &=& -i\eta (p_x-A_x+p_y-A_y).
\end{eqnarray}
\end{subequations}
Putting the Hamiltonian (\ref{HHH3}) into (\ref{VP3}), the force is obtained as
\begin{widetext}
\begin{equation}
\left\langle\mathbf{f}\right\rangle = \frac{\tau_3+i\tau_2}{m_0\hbar}\eta\left[\left\langle p_y-A_y+p_z-A_z\right\rangle\mathbf{i}+\left\langle -p_x+A_x+p_z-A_z\right\rangle\mathbf{j}
   -\left\langle p_x-A_x+p_y-A_y\right\rangle\mathbf{k}\right].
\end{equation}
\end{widetext}

Let us compare the results from two different starting points. The velocity in (\ref{VP2}) is mapped to the Heisenberg representation,
\begin{equation}\label{VP5}
\left\langle\mathbf{\widehat{v}}\right\rangle=\frac{\tau_3+i\tau_2}{m_0\hbar}
\left[\kappa^\alpha\left\langle\mathbf{p}-\mathbf{A}\right\rangle+\kappa^\beta \left\langle \widehat{\mathbf{k}}_v\right\rangle
\right],
\end{equation}
where
\begin{widetext}
\begin{equation}\label{kv2}
\left\langle\widehat{\mathbf{k}}_v\right\rangle= \left\langle p_y-A_y-p_z+A_z)\right\rangle\mathbf{i}+\left\langle p_x-A_x+p_z-A_z\right\rangle\mathbf{j}+\left\langle -p_x+A_x+p_y-A_y\right\rangle\mathbf{k}
\end{equation}
\end{widetext}

It can be seen that the velocity (\ref{VP5}) depends on $\kappa^\alpha$ and $\kappa^\beta$. Note that $\kappa^\alpha=\hbar\left(1+\frac{\theta\eta}{2}\right)$, the first term in (\ref{VP5}) is consistent with the velocity (\ref{VP4}).
This is because the noncommutative relations contain the couplings between coordinates and momenta, but the commutative relation $[\mathbf{x},H]$ in Case II does not contain the couplings between  relations coordinates and momenta.

Interestingly, the forces from these two starting points are same when we transform those of Case I to the Heisenberg representation using the SW map. The free particle carries with an intrinsic velocity and force induced by the noncommutative phase space for both cases.

Consequently, in principle, we seems to have two choices to define oservables. However, the results tell us the definition of observables in the noncommutative phase space contains more noncommutative effects.

\section{Perturbation solution of Klein-Gordon equation}
\subsection{Eigen energies and wave functions}

To explore the basic properties of the KG equation, we solve the KG equation. Note that the noncommutative effects are very small because the noncommutative parameters $\ell_P\ll 1$ and $\Lambda\ll 1$, the KG equation can be solved by using the perturbation theory. The Hamiltonian associated with the KG equation (\ref{HHH1}) can be separated into two parts,
\begin{equation}\label{HH0}
H=H_0+H'
\end{equation}
where
\begin{subequations}\label{HH1}
\begin{eqnarray}
H_0  &=& -\frac{\hbar^2(\tau_3+i\tau_2)}{2m_0}\nabla^2+\tau_3 m_0c^2, \\
H' &=& \frac{\tau_3+i\tau_2}{2m}\left(\frac{\hbar}{i}\mathbf{A}\cdot \nabla + \mathbf{A}^2\right)+A_0,
\end{eqnarray}
\end{subequations}
where $H_0$ is the conventional part in the KG equation and $H'$ is the perturbed part induced by the noncommutative algebra, which can be regarded as a perturbation. Note that
\begin{equation}\label{TTTao}
\tau_3+i\tau_2=\left(
\begin{array}{cc}
1 & 1 \\
-1 & -1
\end{array}\right),
\end{equation}
The Hamiltonian $H_0$ can be expressed as a matrix form
\begin{equation}\label{H01}
H_0=\left(
\begin{array}{cc}
-\frac{\hbar^2}{2m_0}\nabla^2+m_0c^2 & -\frac{\hbar^2}{2m_0}\nabla^2 \\
\frac{\hbar^2}{2m_0}\nabla^2 & \frac{\hbar^2}{2m_0}\nabla^2-m_0c^2
\end{array}\right),
\end{equation}
Suppose that the wave vectors are given by
\begin{equation}\label{H0Wf1}
\psi_p = \frac{1}{(2\pi\hbar)^{3/2}}\left(
\begin{array}{c}
\varphi  \\
\chi
\end{array}\right)
e^{i(\mathbf{p\cdot x}-Et)/\hbar},
\end{equation}
Substituting the wave vectors (\ref{H0Wf1}) into the Schr\"{o}dinger-like equation, $i\hbar \partial_t \psi_p^\pm =H_0\psi_p^\pm$, we have
\begin{equation}\label{SchE1}
\left(
\begin{array}{cc}
\frac{\mathbf{p}^2}{2m_0}+m_0c^2 & \frac{\mathbf{p}^2}{2m_0} \\
-\frac{\mathbf{p}^2}{2m_0} &  \frac{\mathbf{p}^2}{2m}+m_0c^2
\end{array}
\right)\left(
\begin{array}{c}
\varphi \\
\chi
\end{array}
\right)=E\left(
\begin{array}{c}
\varphi \\
\chi
\end{array}
\right)
\end{equation}
Solving the eigen equation (\ref{SchE1}), the eigen energies are obtained\cite{Armin}
\begin{equation}\label{EEg1}
E_\pm=\pm c\sqrt{\mathbf{p}^2+m_0^2 c^2}\equiv\pm cp_0
\end{equation}
where $p_0=\sqrt{\mathbf{p}^2+m_0^2 c^2}$ and the corresponding eigen vectors are given by

\begin{subequations}\label{H0Wf2}
\begin{eqnarray}
\psi^+_p &=& \frac{1}{(2\pi\hbar)^{3/2}}\left(
\begin{array}{c}
m_0c+p_0  \\
m_0c-p_0
\end{array}\right)
e^{i(\mathbf{p\cdot x}-cp_0t)/\hbar} , \\
\psi^-_p &=& \frac{1}{(2\pi\hbar)^{3/2}}\left(
\begin{array}{c}
m_0c-p_0  \\
m_0c+p_0
\end{array}\right)
e^{i(\mathbf{p\cdot x}+cp_0t)/\hbar} ,
\end{eqnarray}
\end{subequations}

Note that the Hamiltonian (\ref{HH1}b) is non-Hermitian, one introduces the generalized inner product, (G-inner product),\cite{Armin} which is defined by
\begin{subequations}
\begin{eqnarray}\label{GINP1}
\left\langle \varphi|\psi\right\rangle_G &:=&\left\langle \varphi|\tau_3|\psi\right\rangle,  \\
\left\langle \varphi|Q|\psi\right\rangle_G &:=&\left\langle \varphi|\tau_3 Q|\psi\right\rangle.
\end{eqnarray}
\end{subequations}

Using the G-inner product, the wave vectors are normalized to
\begin{widetext}
\begin{subequations}\label{H0Wf2}
\begin{eqnarray}
\psi^+_p &=& \frac{1}{2\sqrt{mcp_0}}\left(
\begin{array}{c}
m_0c+p_0  \\
m_0c-p_0
\end{array}\right)
\frac{e^{i(\mathbf{p\cdot x}-cp_0t)/\hbar}}{(2\pi\hbar)^{3/2}}, \\
\psi^-_p &=& \frac{1}{i\sqrt{2(m^2c^2+p_0^2)}}\left(
\begin{array}{c}
m_0c-p_0  \\
m_0c+p_0
\end{array}\right)
\frac{e^{i(\mathbf{p\cdot x}+cp_0t)/\hbar}}{(2\pi\hbar)^{3/2}},
\end{eqnarray}
\end{subequations}
\end{widetext}
Based on the perturbation theory, the 1-order correction of the eigen energyies are given by
\begin{equation}\label{EC1}
E^{(1)}_\pm =\left\langle \psi_\pm|H'|\psi_\pm \right\rangle_G,
\end{equation}
and the 1-order correction of the eigen vectors are expressed as
\begin{subequations}\label{WF2}
\begin{eqnarray}
\psi^{(1)}_{+} &=& \frac{\left\langle \psi_-|H'|\psi_+ \right\rangle_G}{E_+-E_-}\psi_+ \\
\psi^{(1)}_{-} &=& \frac{\left\langle \psi_+|H'|\psi_- \right\rangle_G}{E_--E_+}\psi_- .
\end{eqnarray}
\end{subequations}
Plugging the eigen vectors and the perturbation Hamiltonian into (\ref{EC1}) and (\ref{WF2}), the correction of the eigen energies are obtained
\begin{subequations}
\begin{eqnarray}\label{EC2}
E^{(1)}_+  &=&  \frac{c\mathbf{A}^2+c\mathbf{A\cdot p}+2A_0p_0}{2p_0}\\
E^{(1)}_-  &=&  \frac{cm_0\left(c\mathbf{A}^2+c\mathbf{A\cdot p}\right)-2A_0p_0}{m_0^2c^2+p_0^2}
\end{eqnarray}
\end{subequations}
and corresponding eigen vectors are obtained
\begin{widetext}
\begin{subequations}\label{H0Wf3}
\begin{eqnarray}
\psi^{+(1)}_p &=& -\frac{m_0c\left(\mathbf{A}^2+c\mathbf{A\cdot p}\right)}{4p_0\sqrt{mcp_0}(m_0^2c^2+p_0^2)}\left(
\begin{array}{c}
m_0c-p_0  \\
m_0c+p_0
\end{array}\right)
\frac{e^{i(\mathbf{p\cdot x}-cp_0t)/\hbar}}{(2\pi\hbar)^{3/2}}, \\
\psi^{-(1)}_p &=& \frac{i\left(\mathbf{A}^2+c\mathbf{A\cdot p}\right)}{4p_0^2\sqrt{2(m_0^2c^2+p_0^2)}}\left(
\begin{array}{c}
m_0c+p_0  \\
m_0c-p_0
\end{array}\right)
\frac{e^{i(\mathbf{p\cdot x}+cp_0t)/\hbar}}{(2\pi\hbar)^{3/2}},
\end{eqnarray}
\end{subequations}
\end{widetext}

Thus, we obtain the perturbation solutions of the KG equation in the 1-order approximation,
\begin{equation}\label{EEg3}
E_\pm=\pm cp_0+E^{(1)}_\pm
\end{equation}
and their wave vectors are expressed as
\begin{widetext}
\begin{subequations}\label{H0Wf3}
\begin{eqnarray}
\psi^{+}_p &=& \frac{1}{2\sqrt{mcp_0}}
\left[\left(
\begin{array}{c}
m_0c+p_0  \\
m_0c-p_0
\end{array}\right)-\frac{m_0c\left(\mathbf{A}^2+c\mathbf{A\cdot p}\right)}{2p_0(m_0^2c^2+p_0^2)}
\left(
\begin{array}{c}
m_0c-p_0  \\
m_0c+p_0
\end{array}\right)
\right]
\frac{e^{i(\mathbf{p\cdot x}-cp_0t)/\hbar}}{(2\pi\hbar)^{3/2}}, \\
\psi^{-}_p &=&  \frac{1}{i\sqrt{2(m_0^2c^2+p_0^2)}}\left[\left(
\begin{array}{c}
m_0c-p_0  \\
m_0c+p_0
\end{array}\right)-\frac{\mathbf{A}^2+c\mathbf{A\cdot p}}{4p_0^2}
\left(
\begin{array}{c}
m_0c+p_0  \\
m_0c-p_0
\end{array}\right)\right]
\frac{e^{i(\mathbf{p\cdot x}+cp_0t)/\hbar}}{(2\pi\hbar)^{3/2}}.
\end{eqnarray}
\end{subequations}
\end{widetext}
The physical meanings of these solution are expected to be studied further for some practical issues.

\subsection{Probability and current densities}
Using the wave vectors (\ref{H0Wf3}) and  the probability and current densities (\ref{CD2}) and (\ref{CD3}), we obtain the probability density in the 1-order approximation,
\begin{subequations}\label{CD4}
\begin{eqnarray}
\rho_+  &=&  - \frac{1}{(2\pi\hbar)^{3/2}}\left(1-\frac{m_0c\left(\mathbf{A}^2+c\mathbf{A\cdot p}\right)}{2p_0(m_0^2c^2+p_0^2)}\right)^2 , \\
\rho_-  &=&   \frac{2m_0cp_0}{(2\pi\hbar)^{3/2}(m_0^2c^2+p_0^2)}\left(1-\frac{\mathbf{A}^2+c\mathbf{A\cdot p}}{4p_0^2}\right)^2 .
\end{eqnarray}
\end{subequations}
and the probability current density,
\begin{subequations}\label{JD2}
\begin{eqnarray}
\mathbf{j}_{++}  &=&  \frac{m_0c}{(2\pi\hbar)^{3}}\frac{\left(\mathbf{p}-\mathbf{A}\right)}{p_0}, \\
\mathbf{j}_{--}  &=&  \frac{1}{(2\pi\hbar)^{3}}\frac{2m_0cp_0}{m_0^2c^2+p_0^2}\left(\mathbf{p}-\mathbf{A}\right), \\
\mathbf{j}_{+-}  &=&  -\frac{i\sqrt{2}}{(2\pi\hbar)^{3}}\frac{m^2c^2}{\sqrt{m_0cp_0}}\frac{\left(\mathbf{p}-\mathbf{A}\right)}{\sqrt{m_0^2c^2+p_0^2}}
e^{i2cp_0t/\hbar}, \\
\mathbf{j}_{-+}  &=&  \frac{i\sqrt{2}}{(2\pi\hbar)^{3}}\frac{m_0cp_0}{\sqrt{m_0cp_0}}\frac{\left(\mathbf{p}-\mathbf{A}\right)}{\sqrt{m_0^2c^2+p_0^2}}
e^{-i2cp_0t/\hbar},
\end{eqnarray}
\end{subequations}

Even though the probability density and current density are not positive definite, what observables related to these densities are worth studying further.

\subsection{Nonrelativistic approximation}
In the nonrelativistic approximation, suppose that $|A_0|\ll m_0c$ and $\left|\mathbf{A}\right|\ll m_0c$,\cite{Greiner} we have
\begin{equation}\label{AAA2}
\frac{1}{2m_0}\left(\frac{\hbar}{i}\nabla-\mathbf{A}\right)^2\phi^\pm=m_0c^2\mathcal{O}\left(\frac{v^2}{c^2}\right)\phi^\pm
\end{equation}
and
\begin{eqnarray}
\left(i\hbar\frac{\partial}{\partial t}-A_0\right)\phi^+ &=& m_0c^2\left(1+\mathcal{O}\left(\frac{v^2}{c^2}\right)\right)\phi^+, \\
\left(i\hbar\frac{\partial}{\partial t}-A_0\right)\phi^- &=& m_0c^2\left(-1+\mathcal{O}\left(\frac{v^2}{c^2}\right)\right)\phi^-.
\end{eqnarray}
For the positive solution,
\begin{equation}\label{PS1}
\psi=\left(
\begin{array}{c}
\varphi  \\
\chi
\end{array}\right)=
\left(
\begin{array}{c}
1\\
\mathcal{O}\left(\frac{v^2}{c^2}\right).
\end{array}\right)\varphi.
\end{equation}
The KG equation can be rewritten as
\begin{equation}\label{AAAq}
i\hbar\frac{\partial \varphi}{\partial t}=\left[
\frac{\left(\frac{\hbar}{i}\nabla-\mathbf{A}\right)^2}{2m_0}+m_0c^2+A_0+\mathcal{O}\left(\frac{v^4}{c^4}\right)\right]\varphi.
\end{equation}
For the negative solution,
\begin{equation}\label{NS1}
\psi=\left(
\begin{array}{c}
\varphi  \\
\chi
\end{array}\right)=
\left(
\begin{array}{c}
\mathcal{O}\left(\frac{v^2}{c^2}\right) \\
1
\end{array}\right)\chi
\end{equation}
The KG equation can be rewritten as
\begin{equation}\label{AAAq}
i\hbar\frac{\partial \chi}{\partial t}=\left[
-\frac{\left(\frac{\hbar}{i}\nabla-\mathbf{A}\right)^2}{2m_0}+m_0c^2+A_0+\mathcal{O}\left(\frac{v^4}{c^4}\right)\right]\chi
\end{equation}

The Hamiltonian form of the KG equation can be expressed as
\begin{equation}\label{HKG3}
i\hbar\frac{\partial \psi}{\partial t}=H^{nr}\psi,
\end{equation}
where the Hamiltonian is given by
\begin{equation}\label{NCH1}
H^{nr}=\tau_3\left[
-\frac{\left(\frac{\hbar}{i}\nabla-\mathbf{A}\right)^2}{2m_0}\phi^\pm+m_0c^2+A_0+\mathcal{O}\left(\frac{v^4}{c^4}\right)\right].
\end{equation}
Thus, in the nonrelatativistic approximation, the KG equations (\ref{Hf3}) is decoupled to the positive and negative differential equation.

\section{Symmetry}
Let us explore the fundamental symmetries of the noncommutative relations and the KG equation in the noncommutative phase space.
In the 4D spacetime with the Lorentz symmetry, it can be verified that the noncommutative relation and the KG equation cannot hold invariants under
either parity or time reversal transformation. Thus, we define the composites of the parity and time reversal transformations,

\textbf{Definition}:
\begin{subequations}\label{PT1}
\begin{eqnarray}
\mathcal{PT} & : & \widehat{x}^\mu\rightarrow -\widehat{x}^\mu, \quad  \widehat{p}_\mu\rightarrow -\widehat{p}_\mu, \\
   & \Leftrightarrow & t\rightarrow -t, \quad \widehat{\mathbf{x}}\rightarrow -\widehat{\mathbf{x}}, \quad  \widehat{\mathbf{p}}\rightarrow -\widehat{\mathbf{p}}, \\  && i\rightarrow i.
\end{eqnarray}
\end{subequations}
The second line in (\ref{PT1}b) is equivalent to the first line. Consequently, based on this definition of $\mathcal{PT}$ transformation, $\mathcal{PT}\widehat{x}^\mu\widehat{x}^\nu \mathcal{TP}=\widehat{x}^\mu\widehat{x}^\nu $, we obtain

\textbf{Claim II}: The noncommutative relations are invariant under the parity and time reversal transformations in (\ref{PT1}),
\begin{subequations}\label{PT2}
\begin{eqnarray}
\mathcal{PT}\left[\widehat{x}^\mu,\widehat{x}^\nu \right]\mathcal{TP} &=& \left[\widehat{x}^\mu,\widehat{x}^\nu \right], \\
\mathcal{PT}\left[\widehat{x}^\mu,\widehat{p}_\nu \right]\mathcal{TP} &=& \left[\widehat{x}^\mu,\widehat{p}_\nu \right], \\
\mathcal{PT}\left[\widehat{p}_\mu,\widehat{p}_\nu \right]\mathcal{TP} &=& \left[\widehat{p}_\mu,\widehat{p}_\nu \right].
\end{eqnarray}
\end{subequations}
The angular momentum is defined by $\widehat{L}_\mu:=\varepsilon_{\mu\nu}\ _\kappa\widehat{x}^\nu \widehat{p}_\kappa$ in the noncommutative phase space. {\it Since $\mathcal{PT}\widehat{x}^\mu\mathcal{TP}= -\widehat{x}^\mu,  \mathcal{PT}\widehat{p}_\mu\mathcal{TP}= -\widehat{p}_\mu$,  and $\mathcal{PT}\widehat{L}_\mu\mathcal{TP}=\widehat{L}_\mu$}, thus, we have

\textbf{Claim III}:
\begin{subequations}\label{PT3}
\begin{eqnarray}
\mathcal{PT}\left[\widehat{x}^\mu,\widehat{L}_\nu\right]\mathcal{TP}&=&-\left[\widehat{x}^\mu,\widehat{L}_\nu\right], \\
\mathcal{PT}\left[\widehat{p}_\mu,\widehat{L}_\nu\right]\mathcal{TP}&=&-\left[\widehat{p}_\mu,\widehat{L}_\nu\right], \\
\mathcal{PT}\left[\widehat{L}_\mu,\widehat{L}_\nu \right]\mathcal{TP} &=& \left[\widehat{L}_\mu,\widehat{L}_\nu \right],
\end{eqnarray}
\end{subequations}
Note that the effective gauge potential under the parity and time reversal transformations in (\ref{PT1}) is $A_\mu\rightarrow -A_\mu$,
and we have $\mathcal{PT}\left(\widehat{p}_\mu- A_\mu\right)\mathcal{TP}= -\left(\widehat{p}_\mu-A_\mu\right)$, thus,
we obtain

\textbf{Claim IV}: The canonical form of the KG equation (\ref{KGE1}) is invariant under the parity and time reversal transformations in (\ref{PT1}).

The effective gauge field (\ref{GF2}) is a skew constant matrix. Hence, we have

\textbf{Claim V}: The gauge field $F_{\mu\nu}$ in (\ref{GF2}) is invariant under the parity and time reversal transformations in (\ref{PT1}).

It should be remarked that we should note that the complex variable $i$ does not change under the definition of the time reversal transformations in (\ref{PT1}), which is different from those in the Schr\"{o}dinger equation in the canonical phase space, where $i\rightarrow -i$ . This property could provide some hints for understanding some unsolved puzzles.

\section{Conclusions and outlook}\label{sect7}

The intrinsic spacetime singularities, such as the early universe and black hole, hint the existence of a minimum length in the Planck scale and a minimum curvature of spacetime driving the acceleration expansion of the universe that would be interpreted as dark energy associated with the cosmological constant. These mysterious phenomena lead to many attempts to quantize spacetime background and deform the canonical quantum mechanics.
The noncommutative quantum mechanics shows some novel phenomena in condensed matter physics.\cite{Liang1,Liang2}

We extend the quantization of the 3D to 4D spacetime in the Planck and universe scales. We generalize the noncommutative relations between
the 4-vector operators of position and momentum. Using the Seiberg-Witten (SW) map with the Bopp shift, the noncommutative algebra can be mapped to the Heisenberg commutative algebra. We endow the noncommutative parameters with the Planck constant, Planck length, and cosmological constant. We find that the noncommutative effects can be regarded as an effective gauge field as an analog with the electromagnetic gauge potential. The effective gauge field can be interpreted as the cosmological constant and Planck constant. We give the Klein-Gordon (KG) equation in the noncommutative phase space, including the canonical and Hamiltonian forms of the KG equation. We obtain the current continuity equation. Using the perturbation approach, we obtain the perturbation solution of the KG equation. Moreover, we analyze the fundamental symmetry of the formulation of KG theory in the noncommutative phase space. We find that the noncommutative relations and the KG equation are invariants under the composites of the parity and time reversal transformations if $i$ is invariant.

It should be emphasized that the novel results include
\begin{itemize}
  \item We extend the 3D noncommutative phase space to 4D noncommutative phase space based on the SW map such that we give the KG equation in the noncommutative phase space. We also obtain the perturbation solution of the KG equation and its corresponding probability density and current continuity equation.
  \item We propose a parameterization scheme to endow the noncommutative parameters with the Planck length and cosmological constant such that we can apply this formulation to explore some unsolved puzzles, such as dark energy and intrinsic singularities of spacetime background.
  \item We find that the noncommutative effects can be interpreted as an effective gauge potential. The gauge field depends on the Planck constant and cosmological constant. The gauge field is embedded naturally with the dynamical equation without coupled constant, which is not like the $U(1)$ electromagnetic field and the $SU(2)$ Yang-Mills field. This provides physical scenarios of dark energy and interplay between quantum and gravity.
  \item We find that the free particle carries with an intrinsic velocity and force induced by the noncommutative relations, which could inspire some hints to reveal the physical scenario of the acceleration expending of universe.
\end{itemize}

This formulation opens not only a novel insight into relativistic quantum mechanics in the noncommutative phase space, but also inspires some novel mathematical structures.

\section{Appendix}
\subsection{The basic commutative relations}
For convenience, we compare some basic commutative relations in the Heisenberg phase space and noncommutative phase space.

In the Heisenberg phase space, the basic commutative relations are given by
\begin{subequations}
\begin{eqnarray}  \label{xPP2}
\left[ x,\mathbf{p}^{2} \right]&=& 2i\hbar p_x;\\
\left[ y,\mathbf{p}^{2} \right]&=& 2i\hbar p_y;\\
\left[ z,\mathbf{p}^{2} \right]&=& 2i\hbar p_z,
\end{eqnarray}
\end{subequations}
where $\mathbf{p}^{2}=p_x^2+p_y^2+p_z^2$.

In the noncommutative phase space, using relations of the position and momentum operators in (\ref{XXNCR1}), (\ref{PPNCR1}) and (\ref{XPNCR1}), we have
\begin{subequations}
\begin{eqnarray}  \label{PP2}
\left[ \widehat{x},\widehat{\mathbf{p}}^{2} \right]&=& 2i\left[\kappa^{\alpha}\widehat{p}_x+ \kappa^{\beta}\left(\widehat{p}_y-\widehat{p}_z\right)\right];\\
\left[ \widehat{y},\widehat{\mathbf{p}}^{2} \right]&=& 2i\left[\kappa^{\alpha}\widehat{p}_y+ \kappa^{\beta}\left(\widehat{p}_x+\widehat{p}_z\right)\right];\\
\left[ \widehat{z},\widehat{\mathbf{p}}^{2} \right]&=& 2i\left[\kappa^{\alpha}\widehat{p}_z+ \kappa^{\beta}\left(-\widehat{p}_x+\widehat{p}_y\right)\right].
\end{eqnarray}
\end{subequations}
They can be rewritten as a vector form
\begin{equation}\label{2P3}
\left[ \widehat{\mathbf{x}},\widehat{\mathbf{p}}^{2} \right]= 2i\left(\kappa^{\alpha}\widehat{\mathbf{p}}+\kappa^{\beta}\widehat{\mathbf{k}}_v\right),
\end{equation}
where
\begin{equation}\label{2Pr}
\widehat{\mathbf{k}}_v=\left(\widehat{p}_y-\widehat{p}_z\right)\mathbf{i}+\left(\widehat{p}_x+\widehat{p}_z\right)\mathbf{j}
+\left(-\widehat{p}_x+\widehat{p}_y\right)\mathbf{k}.
\end{equation}
As with the position operator, we have
\begin{subequations}
\begin{eqnarray}  \label{PPpp2}
\left[ \widehat{p}_x,\widehat{\mathbf{p}}^{2} \right]&=& 2i\eta\left(\widehat{p}_y+\widehat{p}_z\right);\\
\left[ \widehat{p}_y,\widehat{\mathbf{p}}^{2} \right]&=& 2i\eta\left(-\widehat{p}_x+\widehat{p}_z\right);\\
\left[ \widehat{p}_z,\widehat{\mathbf{p}}^{2} \right]&=& -2i\eta\left(\widehat{p}_x+\widehat{p}_y\right).
\end{eqnarray}
\end{subequations}
The vector form can be expressed as
\begin{widetext}
\begin{equation}\label{2P3}
\left[ \widehat{\mathbf{p}},\widehat{\mathbf{p}}^{2} \right]=
2i\eta\left[\left(\widehat{p}_y+\widehat{p}_z\right)\mathbf{i}+2i\eta\left(-\widehat{p}_x+\widehat{P}_z\right)\mathbf{j}
-2i\eta\left(\widehat{p}_x+\widehat{p}_y\right)\mathbf{k}\right].
\end{equation}
\end{widetext}
It is quite different from those in the Heisenberg phase space, $[\mathbf{p},\mathbf{p}^2]=0$.
When the noncommutative parameters vanish, the above commutative relations reduce to the canonical relations in the Heisenberg's algebra.

\begin{acknowledgments}
The author thanks the Grant of Scientific and Technological Projection of Guangdong province No: 2021A1515010036.
\end{acknowledgments}


\end{document}